\documentclass[a4paper, 11pt, english]{article}
\usepackage[a4paper,top=2.5cm,bottom=2.5cm,left=2.5cm,right=2.5cm]{geometry}
\usepackage{graphicx} 
\usepackage{epigraph}
\usepackage{authblk}
\usepackage{hyperref}
\usepackage{soul}
\usepackage{xcolor}

\title{Against the “nightmare of a mechanically
determined universe": Why Bohm was never a Bohmian}
\author[1]{Flavio Del Santo}
\affil[1]{\small Group of Applied Physics, University of Geneva, 1211 Geneva, Switzerland; and  Constructor University, Geneva, Switzerland}

\author{Gerd Christian Krizek}
\affil{Department Applied Mathematics and Physics, University of Applied Sciences Technikum Wien, 1200 Vienna, Austria}

\date{\today}

\begin{document}
\bibliographystyle{unsrt}

\maketitle

\begin{abstract}
\noindent David Bohm has put forward the first deterministic interpretation of quantum physics, and for this he seems to be regarded as a champion of determinism by  physicists (both his contemporaries and the supporters of his interpretation, the so-called “Bohmians") as well as by historians of physics. The standard narrative is that he underwent a “conversion" from being a supporter of Bohr to being a staunch determinist, due to his interaction with Einstein and his commitment to Marxism. Here we show that Bohm actually upheld with continuity throughout his career some philosophical tenets that included a strong rejection of mechanistic determinism. As such, we conclude that \textit{Bohm was never a Bohmian} and that his philosophical views have been largely misinterpreted.  
\end{abstract} 

\epigraph{\emph{“Why on earth are they calling it Bohmian mechanics? Haven't they read a word I have written?!"}}{David Bohm (reported by Basil Hiley)}

\section{Introduction}
\noindent  David Bohm (1917-1992) went down in history as the physicist who achieved the impossible by providing an alternative deterministic interpretation of quantum mechanics \cite{bohm1952suggested-1, bohm1952suggested-2}.\footnote{
Bohm himself referred to his interpretation as ``alternative interpretation"\cite{bohm1952suggested-1,bohm1952suggested-2, bohm1957causality}, as ``causal interpretation"\cite{bohm1953proof, bohm1954model}, and as ``quantum potential interpretation". In the literature it is referred to as ``Ontological interpretation" \cite{BohmHiley1993UndividedUniv, pylkkanen2016quantum}, ``De Broglie-Bohm causal interpretation"\cite{holland1995quantum}, or ``De Broglie-Bohm Pilot-Wave Theory", ``Bohmian Mechanics" \cite{durr2009bohmian, maudlin2019philosophy}, or ``Bohm theory" \cite{hiley2001non, esfeld2013ontology}. The variety of terminologies reflects different stances and views of Bohm's collaborators and successors which deviate in some cases substantially from Bohm's own ideas and whose discussion would go beyond the scope of this work.} Acclaimed or blamed therefore as a champion of determinism, he was (and still is) regarded by many as a cure against the claims of the Copenhagen school that quantum mechanics necessarily requires a completely novel way of looking at the world. According to this narrative, Bohm restored the seemingly lost comfort of mechanistic determinism, which had characterized physics for centuries, and his work seems therefore animated by a certain intellectual conservatism (see, e.g.,  \cite{cushing1994quantum}).

Here, we show that it was far from his intention to try to go back to an old pre-quantum paradigm. Bohm's views on philosophy of physics have instead been explicitly aimed, with continuity throughout his whole career, at demolishing certain established views that he perceived as limiting and dogmatic. As we shall see, one of these was the concept of \textit{mechanism}, a form of reductionism which Bohm regarded as the 
\begin{quote}
assumption that the great diversity of things that appear in all of our experience, every day as well as scientific, can all be reduced completely and perfectly to nothing more than consequences of the operation of an absolute and final set of purely quantitative laws determining the behaviour of a few kinds of basic entities or variables. (\cite{bohm1957causality}, p. 37).
\end{quote}
In this effort, Laplacian determinism was regarded by Bohm as the first and foremost expression of mechanism, and he thus searched for alternatives throughout his whole life.

As noted by Nobel laureate Roger Penrose, “there can be few physicists who have delved into the
philosophical implications of their subject as has David Bohm” \cite{Hiley_1997}. It is indeed possible to identify at least three fundamental tenets in David Bohm's philosophy of physics, namely: (i) \textit{realism}, (ii) \textit{causality}, and (iii) \textit{anti-mechanism}. Here we will not deal with Bohm's realism which has already been the subject of numerous studies, and it is undisputed that Bohm was committed to (some form of) realism (see, e.g., \cite{del2018striving, vanStrien2020bohm, junior2019david}, and references therein). On the other hand, we will focus on the latter two tenets, which have been astonishingly misunderstood in most of the vast literature devoted to Bohm's thought and his intellectual legacy. In particular, the term causality %(it ought to be remembered that Bohm himself called his alternative interpretation of quantum theory the ``causal interpretation") 
has been commonly assumed to be a synonym of determinism; a mistake unfortunately still present in the literature in both physics and philosophy to date. Furthermore, Bohm always opposed mechanism, which, we stress again, has its most striking example (but not the only one) in determinism.

It is the main aim of this paper to clarify some of Bohm's original philosophical stances by demolishing certain established misconceptions around his commitment to determinism, which we cannot emphasize enough, was \textit{never} present in his thought. It is a peculiar case that a scholar to whom so many historical and philosophical studies have been devoted has been so misrepresented. Bohm's sustained rejection of determinism was only partly acknowledged in \cite{beller1999quantum} and new important evidences made available thanks to the publication of a collection of letters in  \cite{talbot2017david}. Moreover, one of us (F.D.S.) already pointed out in  \cite{del2018striving} that Bohm's commitment to determinism was secondary to his commitment to realism. The same thesis was then put forward in  \cite{vanStrien2020bohm}. Here, we show that Bohm's position was more radical than this: not only was not determinism his philosophical priority, but he actually always opposed it.

In section \ref{standard}, we will recollect the standard narrative about Bohm's ideas.
%among his contemporaries, historians of physics, and the supporters of the interpretation that went down in history as ``Bohmian mechanics".
Albeit with some variations, indeed, there seems to be a consensus about the fact that Bohm's main philosophical concern was to retrieve determinism in modern physics (at least at a certain stage of his working life). 

We will strongly counter, in section \ref{alternative},  this standard narrative with a more accurate account of the actual philosophical views of David Bohm, focusing on his take on causality and (non)determinism. We will show that one of Bohm's main commitments was always anti-mechanism, a position that he had understood very early to be incompatible with determinism. This is what actually led him to initially (partly) support the indeterministic doctrine of Copenhagen, which, however, he abandoned when he realized that randomness is another, for him unacceptable, form of mechanism.  Hence, his commitment to determinism---stemming from his celebrated alternative interpretation---is only ostensible. Bohm's anti-mechanistic position led him to develop a dialectic philosophical view of an unlimited number of levels of description of reality that can be neither deterministic nor fully random, but still allow either of these descriptions to exist at different levels.

We will here mainly focus on the period of the 1950s, because it is in that decade that Bohm allegedly underwent a change from being a supporter of Bohr to becoming a determinist and then supposedly abandoned this debate altogether as his commitment to Marxism faded away. To avoid further misinterpretations on our part, we will favor quoting as much as possible from Bohm's original writings rather than presenting our own summaries and analyses. Moreover, in the interest of conciseness, but without the risk of decontextualizing the quotations, we will provide more extended excerpts in the form of appendices, where the interested reader can find further evidence in support of the thesis put forward in the main text.  We hope that letting Bohm speak for himself would finally bring justice to some aspects of his complex and original way of conceiving physics.

%%%%%%%%%%%%%%%%%%%%%%%%%%%%%%%%%%%%%%%%%%%%%%%%%%%%%%%%%%%%%%%%%%%%%%%%%%%%%%%%%%%%%%%%%%
\section{The standard narrative: Bohm's alleged commitment to determinism}
\label{standard}

After World War II, the practices of physics underwent a drastic change. The foundational debate that had characterized the early days of quantum physics gave away to a pragmatic approach, the so-called “shut up and calculate", oriented towards applications often of a military nature \cite{kaiser2011hippies}; the debate over the interpretation of the quantum  formalism seemed to be settled for good. It was only a handful of physicists (and a few philosophers) scattered all over the world who started reviving the uneasiness towards the orthodox interpretation proposed by the school of Copenhagen (see Refs. \cite{kaiser2011hippies, junior2014quantum,  baracca2017origins, besson2018interpretation, del2019karl}). Among them, David Bohm was a link between the old generation of critics---such as Albert Einstein, who played and active role in his intellectual life, Erwin Schr\"odinger, or (the early) Luis de Broglie---and the new underground culture concerned with quantum foundations to come.

After completing his PhD with Robert Oppenheimer at Berkeley in the 1940s and a post at the Institute of Advanced Studies in Princeton, in 1951, Bohm fell victim of the witch-hunt of McCarthyism because of his adherence to Marxism; this led him to a life of exile: firstly to Brazil, then to Israel, and finally to the UK, where he spent the rest of his life (see \cite{peat1997infinite, junior2019david} for biographies of Bohm).
 Although his research in the group of Oppenheimer was mainly about plasma physics, it is there that Bohm started getting interested in foundational problems of quantum theory, as he later recalled: “When I went to work with J. Robert Oppenheimer, I found a more congenial spirit in his group. For example, I was introduced to the work of Niels Bohr and this stimulated my interest, especially in the whole question of the oneness of the observer and the observed." (cited in  \cite{junior2019david}, p. 1. See also \cite{peat1997infinite}, Ch. 4). 
 Bohr, together with Werner Heisenberg and others, was not only among the founding fathers of quantum theory but the initiator of the so-called Copenhagen interpretation thereof. The latter maintains that quantum mechanics necessarily 
 leads to abandoning certain fundamental precepts of classical physics, among which determinism, and instead to embrace the genuine probabilistic nature of quantum phenomena.

Bohm went so deep in his reflections about quantum theory and its foundations that, in 1951, he published the textbook 
 \textit{Quantum Theory} \cite{bohm1951quantum}, fully in the spirit of the Copenhagen interpretation. Shortly after the publication, indeed, Bohm himself stated about his book: “a clear presentation of Bohr’s point of view (the first clear, if I may boast a little)."(Letter from Bohm to Miriam Yevick; Letter 66, Folder C117, January 23, 1952. In \cite{talbot2017david}, p. 235.)

However, in the very same year, Bohm submitted, on July 5th, a seminal work  (published in two parts \cite{bohm1952suggested-1, bohm1952suggested-2}) wherein he presented the first consistent alternative interpretation of the quantum formalism. He introduced the initial position of quantum particles as a “hidden variable" that, if known, would lead to \textit{deterministic} trajectories similar to the familiar ones of classical mechanics (but guided by a genuinely additional quantum part in the potential).

So far, these are mere historical facts. Based on these, however, a standard narrative about David Bohm has crystallized, which can be summarized as follows: \textit{In the span of around a year, Bohm had a dramatic shift in his philosophical agenda moving one of his tenets from indeterminism to determinism.} This narrative is not only popularized among physicists in the sort of working history that hovers in the community, but has been advocated by most historians, too. This is however not surprising, since admittedly it \textit{prima facie} seems a rational account of the facts. A more thorough historical reconstruction, proposed among other works in the recent comprehensive biography of Bohm by Olival Freire Jr. \cite{junior2019david}, tells a more nuanced story. First of all, it points out that already in his 1951 book \cite{bohm1951quantum}, Bohm had places some hints of his uneasiness with Copenhagen, such as endorsing ontological realistic assumptions (see \cite{junior2019david}, pp. 48-51). Moreover, historians tend to add a third phase in which Bohm supposedly distanced himself again from determinism at the end of the 1950s, concurrently with his dropping of Marxism. This double shift, also in relation to Marxism, was strongly emphasized already by Pylkk\"anen \cite{pylkkanen1999bohm}, and also Freire, although more cautiously, endorses a similar position:  “Indeed, the connection between the break with Marxism and abandonment of determinism in science, particularly in physics, and not only in society, in Bohm’s thoughts is just a guess, albeit a plausible one." (\cite{junior2019david}, p. 123). At any rate, the main point of the standard narrative is essentially present also in these more informed accounts. 

The historical question that naturally arises then is: \textit{why} did Bohm go through  such a drastic and abrupt change from an adherent of the school of Copenhagen, i.e. a doctrine explicitly advocating the failure of determinism, to a novel deterministic interpretation? (And, possibly, why did he give in determinism again a few years later?). That is, what caused the sudden “conversion" of Bohm from an open supporter of indeterminism to a staunch determinist (and perhaps back)? 

Numerous studies have tried to answer this question (\cite{jammer1974philosophy, peat1997infinite, beller1999quantum,pylkkanen1999bohm,forstner2008early, talbot2017david, junior2019david}, apparently quite successfully despite a few minor details that are still the subject of historical debate.
But what if the question was the wrong one in the first place?  What if determinism has never been a desideratum for Bohm, rather, this change was not about his worldview, but simply it was reflecting different phases of Bohm's experimentation in his attempt to achieve a physical theory that would satisfy his main philosophical tenets? In section \ref{alternative}, we will, in fact, defend this thesis. That is, that Bohm always upheld an anti-mechanistic view that was clearly incompatible with determinism alone.     
Before doing that, in the remainder of this section, we will continue summarizing the standard narrative, or rather, its reply to the main question it poses.

There is an almost absolute consensus on the fact that the two elements that played the major role in Bohm's turn towards determinism have been, on the one hand, his encounter with Einstein, and, on the other, his Marxist views. This twofold explanation is by now well-established among historians, who mostly debate about the extent of one or the other influences (possibly, concurrently with Bohm's political prosecution; see  \cite{forstner2008early}). This reconstruction was already put forward by the illustrious historian and philosopher of physics Max Jammer, according to a late recollection of Bohm himself: 
\begin{quote}
    Stimulated by his discussion with Einstein and influenced by an essay which, as he told the present author, was “written in English” and “probably by Blokhintsev or some other
Russian theorist like Terletzkii,” and which criticized Bohr’s approach, Bohm began to study
the possibility of introducing hidden variables. (\cite{jammer1974philosophy} p. 279)\footnote{Note however, that there is a controversy about the value of this statement because there were no English translations available of either Blokhintsev's or some other Terletzkii's works at the time of Bohm's “conversion". See \cite{junior2019david}, Section 3.4.2.}
\end{quote}
It is indeed well-known that Einstein had opposed Bohr's views since the early days of quantum theory and his attempt to maintain  determinism, summarized by the motto  “God does not play dice", has entered the popular culture. However, while Einstein was invariably troubled by the abandonment of realism (and possibly of locality and localizability) implied by Bohr and his school, there are quite incontrovertible evidences that  determinism was not Einstein's main philosophical concern \cite{del2018striving}, and even less so in his late years. Actually, in 1953, in a letter to his friend Max Born, he stated: “I have written a
little nursery song about physics, which has startled Bohm and de Broglie a little. It is meant to demonstrate the indispensability of your statistical interpretation of quantum mechanics […] This may well have been so contrived by that same ‘non-dice-playing God’ who has caused so much bitter resentment against me, not only amongst the quantum theoreticians but also among the faithful of the Church of the Atheists” (Einstein, A. to Born, M, 12 Oct 1953 \cite{born1971born}). In the light of this, we can conjecture that the impact that Einstein had on Bohm at the time of their encounter at Princeton in the early 1950s, was probably that of casting doubt on the Copenhagen interpretation, and suggesting that one could search for an alternative. However, it does not seem likely that he directly pushed Bohm towards determinism, let alone hidden variable that he never supported (see   \cite{del2018striving}). 

%For what concerns Bohm's  Marxism as the second main influence in developing his deterministic hidden variable interpretation, the question is 

As for whether and to what extent Marxism has been a guiding principle for Bohm in developing his deterministic hidden variable interpretation, the question is subtler. This has been considered in detail by Forstner \cite{forstner2005dialectical, forstner2008early}, and partly by Peat \cite{peat1997infinite}, Freire \cite{junior2019david}, and Talbot \cite{talbot2017david}. Bohm surely agreed with the ontology supported by Marx and Engels, namely, a materialistic philosophy (or \textit{naturalism}) which “says that the sole reality is the natural world, and this world is made up solely of matter" and “material things are not dependent for their existence or nature on any mind or minds", thus implying realism (from A. W. Wood, cited in \cite{talbot2017david}, p. 24). Moreover Marx and Engels put together this materialistic view and the dialectic of Hegel, which turned into the main guiding philosophy of Marxism, i.e., \textit{dialectical materialism}. While dialectical materialism applied in a scientific context deals primarily with the nature of the world, it is in the Marxist analysis of the progress of history and society, \textit{historical materialism}, that one finds determinism as a main characteristic. In fact, for Marx it is the mode of production and the struggle between social classes that \textit{necessarily} determines historical change.

As explained by Freire \cite{junior2019david}, it is objectively difficult to know to which Marxist writings Bohm had access to and therefore which parts of that philosophy had a concrete impact on his scientific and philosophical views. However,  we will see in section \ref{alternative} that it is the dialectic aspect (and partly the materialist one, for what concerns realism) of Marxism that seems to have played the major role in the views about philosophy of science that guided Bohm, rather than the deterministic character of historical materialism.
As a matter of fact, Bohm was already a Marxist when he published his book \cite{bohm1951quantum} in which he endorsed the view of Bohr, so it does not seem to make sense to attribute his alleged conversion towards determinism  to his adherence to Marxism. We will show, on the contrary, that his interest in Bohr actually stemmed, at least partly, from Marxism. This should be regarded as Bohm's first attempt to get away from a mechanistic philosophy in a dialectic (i.e. Marxist) spirit.

Historians are not the only ones who have misconceived Bohm's point of view. The idea that  Bohm's first and foremost concern was that of restoring determinism at any cost was surely always widespread among physicists too. Starting with the contemporaries who were supportive of him---like Einstein, Luis de Broglie, and several Marxist physicists, in particular Jean-Pierre Vigier---and closely followed by his critics, they all emphasized Bohm's commitment to determinism: the former as a merit and the latter as a untenable conservative attitude (see \cite{junior2019david}, Chapters 4.2-4.5, for the early reactions on Bohm's hidden variable model).\footnote{Incidentally, it should be recalled that Bohm's interpretation did not receive the praise that he expected and that he might have deserved. Even Einstein, who supported Bohm in his career and considered him a very talented physicist, stated that the way Bohm's way of restoring determinism “seems too cheap" (see \cite{del2018striving}). There are several hypotheses about why this has been the case, related to the Zeitgeist of post-war physics, Bohm's political views, the authority of the Copenhagen school, etc. (See \cite{junior2014quantum, junior2019david, peat1997infinite, cushing1994quantum}). It was only in more recent years that the so-called Bohmian mechanics found new momentum in a sub-community of scholars interested in foundations of quantum physics (see \cite{durr2009bohmian, durr2012quantum, maudlin2019philosophy}). Also Bohm's close collaborators rediscovered Bohm's original interpretation and encouraged further works closer to Bohm's non-mechanistic ideas (see \cite{baracca1975some}, \cite{philippidis1979quantum}, \cite{BohmHiley1993UndividedUniv}). } As a matter of fact, due to his hidden variable model, Bohm started being regarded as a staunch determinist. 
%Among the supporters of Bohm, the unusual Marxist Léon Rosenfeld wrote to Nobel laureate Frédéric Joliot-Curie to lament that supporters of Bohm in France, such as Jean-Pierre Vigier, “are thinking it is absolutely necessary to kill complementarity and save determinism", 6 Apr 1952; (Freire, p. 85)

%\textbf{[\textcolor{blue}{During the writing I realised that we have to be much more careful about the Bohminas, to avoid destroying a straw man. In general, I think that all the physicists quotations should go in appendix}]}

%%%%%%%%%%%%%%%%%%%%%%%%%%%%%%%%%%%%%%%%%%%%%%%%%%%%%%%%%%%%%%%%%%%%%%%%%%%%%%%%%%%%%%%%%%
%\section{The alternative narrative}
\section{An  alternative narrative: Bohm against mechanistic determinism}
\label{alternative}

\subsection{Indeterminism in Bohm's book \textit{Quantum Theory} (1951) and beyond}
\label{bohm51}
As we have previously recalled, the first work of Bohm in which he manifestly deals with foundational questions is his 1951 book on quantum theory \cite{bohm1951quantum}. It is generally known, as we have discussed, that this book takes an approach close to the orthodox view of Copenhagen. Note that in doing so, Bohm was not blindly following the mainstream, but rather he was actively looking for ways to provide  quantum mechanics of solid and understandable physical foundations, against the wide-spread pragmatic acceptance of an uninterpreted abstract formalism.
%which was tremendously successful in terms of predictive power. 
He therefore saw in the  thought of Bohr an attractive philosophy because it was provided with two main features: the principle of complementarity, and irreducible probability (i.e. nondeterminism). In the former he saw elements of dialectics, which we claim was Bohm's main influence from Marxism. In fact, this is a first attempt, that Bohm was to develop in greater detail in the following years (see below), to apply the ideas of Engels who, in his \textit{Dialectics of Nature}, “is especially opposed to attempts at mechanical
reductionism" \cite{talbot2017david}. In the context of quantum physics, this is the fact that it is the interaction between two qualitatively different descriptions (the classical and the quantum ones) to determine reality, forming something qualitatively new not according to necessity. This also satisfied Bohm's antireductionist convictions because the classical world ought to lie outside of the quantum domain as a primitive and cannot be in general fully reduced to a quantum description. As for the acceptance of objective chance (i.e., potentialities), he saw in this the most natural possibility to abandoning the view of mechanistic determinism. Later Bohm abandoned this approach, but he remained sympathetic to potentialities (see section \ref{potential}). In a letter to at that time his girlfriend Hanna Loewy, presumably in 1950, Bohm explicitly clarified his motivations for having taken a  Bohrian approach in his  book:
\begin{quote}     I just got another idea on the quantum theory also. It is based on the fact that at the microscopic level, the quantum theory deals only with potentialities. For example, the quantum theory describes the probability that an electron can realise its potentiality for a given position. But to realise this potentiality, it must interact with some large scale (classical) system, such as an apparatus which measures position. It is only at the large scale that definite and well-defined events can exist. [...] Thus, the quantum theory presupposes the validity of classical concepts at the classical level. This means that one does not deduce the classical theory from the quantum theory, but that the two work together to describe the whole system. This is in contrast to most theories in physics, in which we analyse all large scale phenomena in terms of the small scale components. Here, we see that at the large scale level, new (classical)
phenomena appear, which are not contained logically in the small scale phenomena
alone. In other words, the behaviour of the whole system cannot be reduced to a
description of the relationship of all its parts, since, new properties appear in a large aggregate, not contained at all in the behaviour of the microscopic systems. (Letter from Bohm to Hanna Loewy; Letter 1. Folder C37, not dated. [February-May, 1950?], \cite{talbot2017david}, p. 99).
\end{quote}
Moreover, soon after the publication of the book, he explained to his friend, the mathematician Miriam Yevick, why he got interested in Bohr:  
\begin{quote}
All I knew was that there was one school, which utterly repelled me, in which one was supposed to introduce abstract mathematical postulates, and be satisfied if the calculations agreed with experiment. Against this, Bohr’s school seemed to be a big improvement, because at least he tried to explain the physical meaning of the theory. Moreover, there was an element of dialectics in Bohr’s point of view which attracted me. It seemed progressive because it broke the old mechanist materialist determinism, which left no room for growth and development of something new. (Bohm to Miriam Yevick; Letter 65. Folder C117, dated: Jan 7, 1952, \cite{talbot2017david}, p. 227); extended quotation in Appendix \ref{yevick23.01.52}).
\end{quote}
Note that at the time when he wrote this letter, Bohm was a staunch Marxist and most remarkably had already completed his work on deterministic hidden variables, and yet he was evidently criticizing \textit{mechanistic materialist determinism}.

For what concerns its content, Bohm's book is an excellent technical manual of quantum mechanics and, although it endorses the view of the Copenhagen school, it is already possible to pin down where the main philosophical concerns of its author lie: causality is already his main focus together with  his refusal of mechanism. However, at this stage, he explicitly endorses indeterminism as a way out of mechanism, a view that was soon to change when he realised that also indeterminism can be mechanistic.

We have recalled in the previous section that Freire \cite{junior2019david} already noticed that a first element that distances Bohm from the Copenhagen school, is that in his 1951 book he looks for a realist account of nature. Another main difference with Copenhagen becomes manifest for what concerns causality. While for Heisenberg ``quantum mechanics proves the invalidity of the law of causality,"\footnote{The original German phrase reads: \textit{``so wird durch die Quantenmechanik die Ungt\"ultigkeit des Kausalgesetzes}".} \cite{heisenberg1985anschaulichen} for Bohm causality was an absolutely indispensable tenet. However, he makes very clear in his book that while maintaining causality he wants to escape determinism. Hence, a first major distinction, surely not well-understood at that time (and alas not even today in most of physics circles), is the conceptual difference between causality and determinism. This is also at the center of misunderstandings in the historical literature when referring to Bohm's  later views, for instance in Freire's words: 
  ``Soon both David Bohm and his critics were using “causal interpretation” to label his approach to quantum theory, clarifying Bohm’s ambition to restore a kind of determinism analogous to that of classical mechanics." (Ref, \cite{junior2019david}, p. 63). In his 1951 book, Bohm actually advocates a causally non-deterministic nature of physical laws, in terms of tendencies (as we will see later, this is closely related to Popper's view in terms of \textit{propensities;}  see section \ref{potential}): 
\begin{quote}
    we wish to call attention to the fact that, even in very early times, two alternative general types of causal laws appeared. One of these involved the notion of complete determinism; the other involved the notion of causes as determining general tendencies but not determining the behavior of a system completely. (\cite{bohm1951quantum}, Ch. 8, Sect. ``Completely Deterministic vs. Causal Laws as Tendencies.")
\end{quote}
Bohm goes as far as to brilliantly show that actually the determinism of classical physics makes the concept of causality redundant: 
\begin{quote}
    It is a curiously ironical development of history that, at the moment causal laws obtained
an exact expression in the form of Newton's equations of motion, the idea
of forces as causes of events became unnecessary and almost meaningless.
The latter idea lost so much of its significance because\textit{ both the past and
the future} of the entire system are determined completely by the equations
of motion of all the particles, coupled with their positions and
velocities at any one instant of time. Thus, we can no more say that
the future is caused by the past than we can say that the past is caused
by the future. [...] 

%Of course, the notion of forces as causes can be retained; in fact, this procedure appears to be the most convenient one to use in practice. From a purely logical point of view, however, the concept of force is redundant, because it is always possible, in principle, to express all classical physics in terms of the positions, velocities, and accelerations of all the particles in the universe. [...]

Thus, classical theory leads to a point of view that is prescriptive and not causal.
(\cite{bohm1951quantum}, Ch. 8, Sect.  ``Classical Theory Prescriptive and not Causal".)
\end{quote}
Hence, he saw a way out of the effective lack of causality in a completely deterministic theory in terms of the tendencies or potentialities entailed by (the Copenhagen interpretation of) quantum physics:
\begin{quote}
With the advent of quantum theory, the idea of complete determinism was shown to be wrong and was replaced by the idea that causes determine only a statistical trend, so that a given cause must be thought of as producing only a tendency toward an effect. [...] (\cite{bohm1951quantum}, Ch. 8, Sect.  ``New Properties of Quantum Concepts : Approximate and Statistical Causality".)

Thus, in terms of our new concept, matter should be regarded as having potentialities for developing either comparatively well-defined causal relationships between
comparatively poorly defined events or comparatively poorly defined
causal relationships between comparatively well-defined events, but not both together. (\cite{bohm1951quantum}, Ch. 8, Sect. ``Relation between Space Time and Causal Aspects of Matter".)
\end{quote}

We have thus seen why Bohm became aligned with Bohr in the first place, namely, to find a suitable alternative to mechanistic determinism that precluded a sensible concept of causality, which was for Bohm a crucial assumption for a  physical theory. However, he soon realized that Bohr’s philosophy was not as satisfactorily as he previously had sensed because it indeed contained a dialectical approach but not as much of materialism as he would have wanted: 
\begin{quote}
    After I had written the book, I finally began to grasp the full
meaning of the theory, and could see that it leads inevitably to a form of (dialectical)
idealism. But this was not so clear when I started, because of the general confusion
in the literature. 
(Bohm to Miriam Yevick; Letter 65. Folder C117, dated: Jan 7, 1952, \cite{talbot2017david}, p. 227); extended quotation in Appendix \ref{yevick23.01.52}).
\end{quote}
And again:
\begin{quote}
I notice that you call me ``a disciple of Einstein". This is not very accurate. Actually I was a strong ``Bohrian" and wrote my book under the assumption (later proved wrong) that the principle of Complementarity was a materialist point of view. It certainly is very dialectical, but I did not see at that time that it is not materialist. 
After writing my book, I sent a copy to Einstein. He called me up asking to discuss the book, especially the Section on the paradox of EPR, which he liked very much. He thought I gave Bohr's point of view the most convincingly possible presentation, but he still refused to accept it. He then argued for some time, and he ended up convincing me that his objections were not answered. I thought about it for a while, becoming more convinced all the time that he was right. Finally I decided to look for a causal interpretation within few weeks, I hit upon the idea which I published, not knowing about de Broglie's work until later. It took me 10 hours of work, distributed over 2 months to convince Einstein that it made sense, but he actually never liked it. He only thought it was good to propose it to break out the present stagnant situation in physics. 
(Bohm to Schatzman; Letter A1.15. September 7, 1952, \cite{besson2018interpretation}, p.335)
\end{quote}

\subsection{Against determinism, despite hidden variables (1952)}
Exactly in the same period when his book \cite{bohm1951quantum} was appearing, Bohm was formulating his alternative, deterministic interpretation in terms of hidden variables.
Given his clear motivation recalled in the previous section, why did he do that? Bohm must have found himself in a strange position, when he managed to conceive a consistent model based on hidden variables that restored determinism. He clearly wanted to prove something that was considered impossible by the founding fathers of theory, in particular John von Neumann who had allegedly proven that a hidden variable completion of quantum mechanics was in principle impossible.\footnote{On the history of von Neumann's impossibility proof see \cite{dieks2017neumann}.} Moreover, Bohm wanted to prove that Bohr and Heisenberg's view was not necessarily the ultimate description of reality.
It should be stressed that at that time, no other interpretation of quantum physics was known besides (slightly different understandings) of the Copenhagen one, so probably stimulated by his novel awareness of the limits of Bohr's interpretation and by the discussions with Einstein he explicitly looked for \textit{any} alternative different interpretation. 
%Bohm expressed this clearly, although much later:
%\begin{quote}
%    To show that it was wrong to throw out hidden variables
%because they could not be imagined, it was therefore sufficient
%to propose any logically consistent theory that explained the
%quantum mechanics, through hidden variables, no matter how
%abstract and hypothetical it might be. (\cite{bohm1980wholeness}, p. 104).
%\end{quote}
According to Hiley, indeed, Bohm
``was not a deterministic man, he used causality. [...] He was not bound to it [determinism]. David Bohm always used to say to me: `I am making a proposal'. So, all this people think he had rigid views. He didn't have rigid views. He was always making proposals, because he thought he never fully got to the bottom of quantum mechanics." \cite{DelSantoKrizek2019}.
    %[...] So then sometimes he proposes thinks which looks like he is a rigid determinist... he is not, he is just exploring ``how far can I take it?", and still find coherence. And the stochastic was another thing, ``let's try".

In fact, although Bohm stresses in his papers that the ```hidden" variables determine the precise results of each individual measurement process" \cite{bohm1952suggested-2}, repeatedly acknowledging very clearly the deterministic character of his model, he certainly never adopted a fundamental ontology merely made of particles plus their deterministic dynamics guided by the wave function. This is something that his followers, the so-called Bohmians (see footnote 1), have instead assumed, namely, considering Bohm's proposal as the ultimate description of reality, much against the view of Bohm himself. In fact, the germ of Bohm's way out of mechanical determinism (see further) as entailed by his proposal, is already expressed, although quite subtly, already in the conclusion of his second paper on hidden variables \cite{bohm1952suggested-2}, when he states:
\begin{quote}
%Finally, as an alternative to the positivist hypothesis of assigning reality only to that which we can now observe, we wish to prevent here another hypothesis, which we believe corresponds more closely to conclusions that can be drawn from general experience in actual scientific research.
This hypothesis is based on
the simple assumption that the world as a whole is
objectively real and that, as far as we now know, it can
correctly be regarded as having a precisely describable
and analyzable structure of unlimited complexity. The
pattern of this structure seems to be rejected completely
but indirectly at every level [...].
%, so that from experiments done at the level of size of human beings', it is very probably possible ultimately to draw inferences concerning the properties of the whole structure at all levels. 
We should never expect to obtain a complete
theory of this structure, because there are almost
certainly more elements in existence than we possibly
can be aware of at any particular stage of scientific
development. Any specified element, however, can in
principle ultimately be discovered, but never all of
them.
% Of course, we must avoid postulating a new element for each new phenomenon. But an equally serious mistake is to admit into the theory only those elements which can now be observed.
\end{quote}

Indeed, at least since 1951, most likely when he was still in Princeton (see \cite{talbot2017david}, footnote 48, p. 31), Bohm started developing a new philosophy based on the concept of having different levels of description, each of which can be either deterministic or indeterministic, but each of them giving only a partial account of reality. His ontology was thus made of the wholeness of the different levels of qualitatively different entities. However, he postulated the number of levels to be infinite, thereby making it fundamentally impossible to have mechanism, and in particular determinism:
\begin{quote}
    Because of the existence of an infinite number of levels, the deterministic laws
of order at each level probably follow only as a result of conditions of chaos existing
at lower levels. If the lower-level conditions of chaos could be altered, then the very
framework of description of the higher level laws would also have to be altered.
Thus, we are led to a more dynamic concept of the laws of nature; for because
of their infinite complexity, richness, and depth, the applicability even of certain
very general forms of laws at a particular level may depend on conditions at other
levels, which are in principle subject to our prediction and control. This experience
should ultimately be repeated at any given level, however deep, as our knowledge is
extended. (Bohm to Miriam Yevick; Letter 58. Folder C116, dated: Nov 23 [1951], \cite{talbot2017david}, p. 205)
\end{quote}

    Note that this idea, while keeping being refined, remained essentially unchanged throughout Bohm's transition from the period of his 1951 book to his hidden variable proposal, and reached its main expression in the book \textit{Causality and Chance} \cite{bohm1957causality} published in 1957 (see section \ref{causal}). For instance, after he had already completed his hidden variable interpretation, he wrote to Yevick:    
    \begin{quote}The “things” at each level, are made up of
smaller “elements” at a more fundamental level, and it is the  motion of these more fundamental
elements (not usually directly visible to us, except with the aid of elaborate
scientific research) which causes the appearance and disappearance of the “things”
existing at a higher level. These more fundamental “elements” however, cannot be
permanent, but must be made up of still more fundamental “elements” and so on ad
infinitum. 
(Bohm to Miriam Yevick; Letter 65. Folder C117, dated: Jan 7, 1952, \cite{talbot2017david}, p. 227; extended quotation in Appendix \ref{yevick07.01.52})
\end{quote}

%put at the right position somewhere here.
Bohm also points out his position on the need for infinite levels to this collaborator Schatzman in a letter from 1952:

\begin{quote}
It is most likely that not even the substratum particles could be indestructible and unanalysable. Instead, there is probably another
substratum below this (of a qualitatively different kind most probably) and so on ad
infinitum. Thus, we should have an infinite series of qualitatively different levels of
laws. Any finite number of levels can always be understood by humanity, but never
\underline{all} of them. (\cite{besson2018interpretation}, p. 351; extended quotation in Appendix \ref{schatzman1})   
\end{quote}

And soon after his letter to Miriam Yevick in January, he wrote what is one of the most important quotations from the whole collection of known writings of David Bohm, because it unambiguously states that he could not accept mechanic determinism, even in the period when he was promoting his hidden variable model:
\begin{quote}
    Most of the errors of both
the positivist and the 19th century “mechanical” materialists spring from an implicit
assumption that the laws of nature will some day finally be understood in terms
of a limited number of hypotheses. From this comes the nightmare of a mechanically
determined universe that follows an inevitable course. To avoid this nightmare,
positivists and idealists have given up causality and assumed a “spontaneous” (i.e.,
uncaused) element in physical processes.

The concept of a limitless number of levels [...] provides a motive
power for continual development \& growth. Moreover, the nightmare of complete
determinism is avoided. Although each level is causal, the totality of levels cannot
ever be taken into account. Thus, as a matter of principle, we say that complete determinism
could not even be conceived of, yet, each level can be determined. Here, we
part company with the believers in “spontaneity” for we say that what appears to be
spontaneous is caused by factors, in principle, knowable, but now hidden to us. But
to be able to say this without implying complete determinism, we must assume an
unlimited number of levels. 
(Bohm to Miriam Yevick; Letter 73. Folder C118, dated: Rec Mar 31 [1952], \cite{talbot2017david}, pp. 254-55; extended quotation in Appendix \ref{yevick31.03.52})
\end{quote}

It is now clear that Bohm did not undergo a conversion form indeterminism (à al Copenhagen) to determinism (with hidden variables), as the standard narrative implies. He actually stayed faithful to his tenets of realism and causality and his shift was merely that of realising that Bohr`s approach was not enough to achieve what he had in mind. So it seems that his philosophical theory of the infinite levels was conceived to “cure" his own model from the “nightmare" of determinism. One should also remark that this idea of unlimited levels is very much in the spirit of dialectics, and indeed this is the most Marxist trait in Bohm's work. As  pointed out by Talbot, such a connection is perhaps less abstract that one could think, drawing directly from the work of Engels:  ``especially in the
Dialectics of Nature, Engels introduces the idea of levels, or what he calls `forms of motion'.  [...] Engels is especially opposed to attempts at mechanical reductionism, which `blots out the specific character' and `qualitative difference' of
non-mechanistic forms of motion." ( \cite{talbot2017david}, p. 25).
For Bohm this dialectic view of nature is a way to maintain a non trivial form of causality, intended as the possibility of creating non necessary new things, contrarily to the mechanistic view. In a letter to his friend ---the American physicist  Melba Phillips--- Bohm spelled out this connection in detail: 
\begin{quote}
Also an important additional aspect of causality needs to be discussed in more detail ---namely--- causality as a means of determining the mode of being of qualitatively
new things, which grow out of the old things. The basic aspect of mechanism is that
(as in an idealized machine) the universe is conceived of as made of basic elements
(particles, fields, or what have you) which simply interact according to fixed roles, and
which themselves never change as a result of the processes in which they take part. [...]  However, the concept of the infinity of levels shows
that there need exist in nature no such thing as a basic element which never changes.
Thus, causal laws not only determine the future in a mechanical sense; i.e., in the
sense of determining quantitative changes in the arrangements of entities whose
intrinsic character is fixed. The causal laws also tell when qualitative changes will
occur and may define the characteristics of the new entities that can come into being. Thus, causality is a broader concept than that of mechanical determinism. [...]
A “mechanistic” attitude toward science however, tends
to limit the growth of our concepts in an arbitrary and dogmatically conceived way.
Such a mechanistic attitude refers not only, however, to the mechanistic determinists,
but also to the “mechanistic indeterminists”, who insist that in the quantum of action, we have reached an ultimate, indivisible, and unanalyzable entity, which will never be found to have a structure understandable in terms of a deeper level.
%In fact, the quantum of action presents many aspects of the ultimate particles of the atomists, so that the insistence that the quantum will never be analyzed is as mechanistic as a  theory of point particles following determined orbits. Similarly, the insistence that chance+probability are not subject to a causal analysis at a deeper level constitutes a mechanistic attitude toward these things, since chance+probability are conceived of as existing in themselves and functioning under all possible circumstances according 
to fixed rules.
    (Bohm to Melba Phillips. Letter 43. Folder C48, dated: Oct 13, 1953, \cite{talbot2017david}, p. 164; extended quotation in Appendix \ref{phillis13.10.53}).
\end{quote}
In the following years, Bohm kept developing his  philosophy of the infinite levels, sharpening the distinction between causality and deterministic mechanism, advocating the former and in strong opposition to the latter. Causality is for Bohm the possibility of creating new qualitative entities in a non trivial sense, i.e. without being able to reduce everything to a finite collection of basic elements that cannot change and that are subject to fix laws:
\begin{quote}
    Now, at first sight, it
may seem that we could eliminate the large-scale level by analyzing it in terms of its
basic molecular motions. And if there were a finite number of levels, this would be
true. But if there are an infinite number, then each level stands on a footing that is, in the long run, as basic as that of any other. For every level has below it a deeper one. Indeed, matter can be regarded as made up of the totality of all levels. Each level
makes its own specific contribution to the totality.  (Bohm to Melba Phillips. Letter 46. Folder C48, dated: March 15, 1954, \cite{talbot2017david}, p. 170; extended quotation in Appendix\ref{phillis15.03.54}).
\end{quote}

Let us now stop for a moment and go back to the standard narrative. Freire makes a case that 

\begin{quote}
in the 1950s Bohm did indeed promote the recovery of determinism. In 1951, before the term `causal interpretation' had gained currency in the debates on Bohm’s proposal, he himself emphasized it in his first letter to the
French astrophysicist and Marxist Évry Schatzman, while looking for allies, such
as Jean-Pierre Vigier and Louis de Broglie, to get support for his proposal: “My
position in these physical questions is that the world along with all observers who
are part of it is objectively real and in principle precisely definable (with arbitrarily
high accuracy), and subject to precise causal laws that apply in each individual case
and not only statistically.” (\cite{junior2019david}, p. 65).     
\end{quote}
There seems to be a tension between the statements of Bohm here. However, one can hypothesize that his actual point of view on determinism is the one that emerges from the letters to his intimate friends, i.e., a staunch anti-mechanistic position. Thus, these letters seem to be a more trustable source than a first contact to somebody from whom Bohm was seeking the support. He probably tamed his more complex philosophical positions and tailored his letters to his interlocutor by highlighting the deterministic aspect in the interactions with Schatzman and later with Vigier to find a common ground with these more ``traditional" Marxists who definitely prised determinism (see Appendix \ref{freire91}). Moreover, note that in the quoted letter to Schatzman, Bohm stresses the causal aspect of his proposal, which, as clarified above, does not necessarily means determinism.

\subsection{An indeterministic causal model by Bohm and Vigier (1954)}
So far, the evidence that Bohm was against determinism even during the years in which he devised and promoted his hidden variable model are limited to private correspondence. However, in 1954, Bohm published a paper with Vigier---\textit{Model of the causal interpretation of quantum theory
in terms of a fluid with irregular fluctuations} \cite{bohm1954model}---that is a first attempt to put into practice the ideas of a model of causal interpretation which is however fundamentally  non-deterministic, due to different levels of description. In fact, therein Bohm and Vigier postulate a field that is described by a fluid of density $|\psi|^2$, which is then able to recover the standard quantum mechanics \begin{quote}
    by introducing the hypothesis of a very irregular and effectively random fluctuation in the motions of the fluid. [...] Such random fluctuations are evidently consistent within the framework of the
causal interpretation of the quantum theory. Thus,
there are always random perturbations of any quantum
mechanical system which arise outside that system. \cite{bohm1954model}
\end{quote}
They indeed clarify that ``the causal interpretation of the quantum theory permits an unlimited number of new physical models" and that their proposed ``model is an extension of the causal interpretation of the quantum theory already proposed, which provides a more concrete physical image of the meaning of our postulates than has been available before, and which suggests new properties of matter that may exist at deeper levels." \cite{bohm1954model}. Here causal means the possibility of explaining the theory in terms of a sub-quantum level (the fluid) that accounts for the higher quantum level. Note that, contrarily to the first hidden variable model \cite{bohm1952suggested-1, bohm1952suggested-2}, this model is based on fundamental random fluctuations, thereby dispelling even more the doubt that Bohm was a committed determinist: “In the model that we have proposed here, however, the statistical fluctuation in the results of such [quantum] measurements are shown to be ascribable consistently to an assumed deeper level of irregular motion”. It is interesting to notice that while the postulated fluctuations of the fluid are considered to be (at this level of description) genuinely indeterministic, Bohm and Vigier think of these fluctuation as having a certain structure in terms of potentialities: “The fact that the mean density remains equal to $|\psi|^2$, despite the effects of the random fluctuations, implies then that a systematic tendency must exist for fluid elements to move toward regions of high mean fluid density.”
The ontological basis of this new indeterministic model and how it relates to Bohm’s philosophy of the infinite levels is explained by Bohm in correspondence with Einstein:
\begin{quote}
 “The general idea is that at a level more fundamental than that of quantum mechanics, there is a field which satisfies causal laws. This field is, however, in a state of statistical fluctuations. These fluctuations are somehow described by the $\Psi$ field.” (Bohm to Einstein ; Letter 16. page 5 Folder C14, February 3, 1954, \cite{BohmletterC14toEinstein}, p. 5).

My own point of view is that below the quantum theory there exists a sub quantum-mechanical level of continuous and causally determined motion, and that the quantum theory is related to the sub-quantum mechanical level, more or less as ordinary Brownian motion is related to the atomic level. 
In other words, events at the atomic level are contingent on the (in general irregular) motions of some as yet unknown but qualitatively new kind of entity, existing below the atomic level. 
As a result, the relationships between things, that can be defined at the atomic level will be characterized by the laws of chance, since they will be determined only in terms of some quasi-ergodic type of motion of new kinds of entities existing at the lower level. (Bohm to Einstein; Letter 21. Folder C15, dated: November 14, 1954, \cite{BohmletterC15toEinstein})
\end{quote}

Einstein’s replies may seem surprising to those who still believe that he was also a committed determinist at any cost, because they show once more that he was dissatisfied with Bohm’s first (deterministic) hidden variable model: “I am glad that you are deeply immersed seeking an objective description of the phenomena and that you feel that the task is much more difficult as you felt hitherto.” (Einstein to Bohm ; Letter 17. Folder C14, February 10, 1954, \cite{BohmletterC14toEinstein}). And again: “In the last years several attempts have been made to complete quantum theory as you have also attempted. But it seems to me we are still quite remote from a satisfactory solution of the problem.” (Einstein to Bohm ; Letter 20. Folder C15, dated: October 28, 1954, \cite{BohmletterC15toEinstein})

Bohm did not develop further this approach which he most likely perceived as well as a proposed first step towards his philosophy of levels of description, but he came back to a stochastic causal interpretation, also with Hiley, in the 1980s \cite{bohm1988non, bohm1989non}.

\subsection{Causality and Chance in Modern Physics (1957)}
\label{causal}

%\textcolor{blue}{[\textbf{This subsection is very important but I have not yet found the time to go through it. So far only a collection of notes. All the relevant quotations from \cite{bohm1957causality} are already pasted below.} ]}

It is around the same period that Bohm started thinking not only that either a deterministic or an indeterministic description was possible in every level of an infinite series, but that both individual laws and statistical laws are necessary for a causal interpretation:
\begin{quote}
   The picture which I propose is this: The totality of causal laws includes both statistical and individual laws. We start with this totality as our basic reality. [...] The fundamental reality is that of matter in being and in process of change, or of becoming, as it may more accurately be called. (Bohm to Miriam Yevick. Letter 121. Folder C124, dated: Sept 10 1954, \cite{talbot2017david}, p. 419-22).

\end{quote}
These dialectic ideas grew into a book, \textit{Causality and Chance}, which Bohm published in 1957 \cite{bohm1957causality}.  Therein, Bohm identifies two types of causal laws (both considered fundamental): simple causal laws that connect past and future one-to-one (i.e. deterministic), and more general ones that are one-to-many, (i.e. that do not lead to a unique evolution but only to an array of possibility):
\begin{quote}
    [L]et us note that the one-to-many character of a causal law has no essential relationship
to a lack of knowledge on our part concerning the additional causal factors to which the more precise details
of the effect can be traced. [...] In other words, a one-to-many law
represents an objectively necessary causal connection, but in this case, what is necessary is that the effect
remain within certain bounds; and not, as in simpler types of causal laws, that the effect be determined
uniquely. (\cite{bohm1957causality}, p. 17).
\end{quote}
And again, Bohm clarifies, as he always maintained (cf. \ref{bohm51}), that causality is a more general concept than that of necessity (i.e., determinism):
\begin{quote}
    We see, then, that it is appropriate to speak about objectively valid laws of chance, which tell us about a side of nature that is not treated completely by the causal laws alone. Indeed, the laws of chance are just as
necessary as the causal laws themselves. [Footnote:] Thus necessity is not to be identified with causality, but is instead a wide category. (\cite{bohm1957causality}, p. 23).
\end{quote}
Furthermore, Bohm here again stresses the fact that objective chance should be interpreted, as a potentiality, i.e., a property of the system and its causal conditions: 
\begin{quote}  
On the basis of the above considerations, we are then led to interpret the probability of, for example, a
given result in the game of dice as an objective property associated with the dice that are being used and
with the process by which they are thrown (\cite{bohm1957causality}, p. 27; extended quotation in Appendix \ref{causality1})
\end{quote}
Note that this example is exactly the same used by Karl Popper \cite{popper1959propensity} when he introduced the propensities interpretation (see section \ref{potential}), again showing the compatibility between Bohm and a worldview based both on causality and on indeterminism.

Beyond causality, a large part of Bohm's 1957 book \cite{bohm1957causality} is devoted to defend another of his  main tenets, namely, anti-mechanism. However, while being still convinced that determinism is an unacceptable form of mechanism, there is a fundamental difference with respect to his book on quantum theory \cite{bohm1951quantum}. Here, in fact, Bohm does not consider randomness alone as a way out of mechanism:
\begin{quote}
    The point of view described above evidently renounces an important aspect of the various forms of the
mechanistic philosophy that appeared from the sixteenth through the nineteenth centuries; namely, their
determinism. But in doing this, it has conserved and in fact enhanced the central and most essential
characteristic of this philosophy; namely, the assumption that everything in the whole universe can be
reduced completely and perfectly to nothing more than the effects of a set of mechanical parameters
undergoing purely quantitative changes. [...] 

The question of what constitutes a mechanistic philosophy, therefore, cuts across the
problems of determinism and indeterminism. For this reason, we shall call the philosophy described in this
section by the name of “indeterministic mechanism” (\cite{bohm1957causality}, pp.62-63).
\end{quote}

Bohm's  criticism of mechanism (and thereby of determinism), does not spare his own hidden variable interpretation, which he considers again an unsatisfactory physical model, whose main feature, he stresses, is consistency:
\begin{quote}
While our theory can be extended formally in a logically consistent way by introducing the concept of a wave in a 3N-dimensional space, it is evident that this procedure is not really acceptable in a physical theory, and should at least be regarded as an artifice that one uses provisionally until one obtains a better theory in which everything is expressed once more in ordinary three-dimensional space. (\cite{bohm1957causality}, p. 117)
\end{quote}

Finally, in his \textit{Causality and Chance}, Bohm for the first time defends publicly his philosophical view of the infinite levels of description as the main alternative to mechanism, be it deterministic or indeterministic (see Appendix \ref{causality1} for relevant quotations). As noted already by Freire \cite{junior2019david}, this marks Bohm's entry in the philosophical debate and would allow him to engage with prominent philosophers of science, the like of Paul Feyerabend and Karl Popper (see further). However, these ideas of infinite levels were not appreciated by his more traditional Marxist followers, who saw in this the undermining of determinism: a positive feature for Bohm and an unacceptable price for them. This is the case of Évry Schatzman and and Vigier who wrote to Bohm: “We may be wrong, but we do not agree at all with your ideas about the different levels of reality. It seems to us that it is a formal interpretation of the famous sentence of Lenin, in Materialism and Empiriocriticism, about the different levels of reality” (quoted in \cite{junior2019david}, p. 108).

To conclude, in \textit{Causality and Chance} Bohm synthesizes his main philosophical tenets that have been present in his writing since the beginning, but in a quite scattered  way. Therein, Bohm defends, for the first time systematically,   causality in its broadest sense, advocating the fundamental necessity of both individual laws and statistical laws, depending on the context. Moreover, he firmly rejects mechanism, not only in the form of determinism (which he did for many years already), but also in its indeterministic form. Finally, Bohm opposes mechanism with a dialectic philosophy of infinite levels of description that he had developed throughout the 1950s.

For what concerns physics proper, in 1957, Bohm published with his student Yakir Aharonov a paper where he rejects his own 1952 model, not on the  basis of determinism but on nonlocality: “It must be admitted, however, that this quantum potential seems rather artificial in form [...] that
it implies instantaneous interactions between distant particles, so that it is not consistent with the theory of relativity.” \cite{PhysRev.108.1070}.  Bohm thus kept proposing his dialectical views of different levels, similar to the paper with Vigier  \cite{bohm1954model}, looking for a a “deeper subquantum-mechanical level” \cite{PhysRev.108.1070}.

%\begin{quote}
%    The singularity of Bohm’s views in this book,
%given his defense of the causal interpretation, was that causal laws and statistical laws appear to have the same epistemological status. Thus Bohm was weakening the prominence he had previously attributed to causal laws in science. The other idea present in the book is the infinity of levels of reality. Bohm expressed this idea in his
%letters as early as November 1951 and his recollections suggest he may have been inspired by the idea of the inexhaustibility of the electron, which he picked up from Lenin’sworks while still at Princeton. This idea had acquired a concrete physical form for him since him and Vigier had modeled the hidden-variables interpretation based
%on the existence of a sub-quantum level of reality which had random movements.72
%Bohm’s independent thinking is evident when we note that these new ideas were not initially shared by all his fellow Marxist physicists. Évry Schatzman, speaking on behalf of Vigier and himself, wrote to Bohm “We may be wrong, but we do not agree at all with your ideas about the different levels of reality. It seems to us that it is a formal interpretation of the famous sentence of Lenin, in Materialism and
%Empiriocriticism, about the different levels of reality”
%\end{quote}

%Wrong! He makes it clear that causality is the main thing, but not determinism.

It is interesting to notice, that still at this stage, Bohm's views were completely misunderstood. Luis de Broglie, who wrote the forward of his Causality and Chance, for instance, keeps attributing to Bohm the great merit of giving hope to those who look for a deterministic hidden variable explanation of quantum theory: “It is possible that looking into the future to a deeper level of physical reality we will be able to interpret the laws of probability and quantum physics as being the statistical results of the development of completely determined values of variables which are at present hidden from us. It may be that the powerful means we are beginning to use to break up the structure of the nucleus and to make new particles appear will give us one day a direct knowledge which we do not now have of this deeper level." (\cite{bohm1957causality}, p. x). This goes completely against what Bohm conveys in his book, making wander whether people like de Broglie were actually reading Bohm’s works or they just imposed on him what they wished to hear. 

Towards the end of the 1950s Bohm abandoned Communism, following the revelations of Stalin’s crimes by Nikita Khrushchev in 1956 (see \cite{junior2019david}). As already recalled, this has been identified in the literature as the main motivation to abandon his commitment to determinism. But as we have shown, such an alleged commitment to determinism was never present in the first place and his dialectic attitude remained an important factor in his philosophy. However, probably due the frustration of being continuously misunderstood, Bohm’s engagement with different models of the causal interpretation became sparser. Actually, since his moving to the UK, firstly in Bristol and then in London, he engaged more and more in the philosophical debate, becoming friend with Paul Feyerabend, Karl Popper and Stephen K\"orner, and he kept his interpretational considerations away from his physics colleagues. 

Hiley joined Bohm at Birkbeck college in London in 1961 and, as a matter of fact, they passed ``ten years without actually talking about the causal interpretation" \cite{DelSantoKrizek2019}. As recalled by Hiley \cite{DelSantoKrizek2019}, it was only in the 1970s that two of Bohm's students, Chris Dewdney and Chris Philippidis, ``rediscovered" the hidden variable papers \cite{bohm1952suggested-1, bohm1952suggested-2} and went to Hiley to ask why Bohm and him were not discussing this important results. Hiley replied ``because it is all wrong", but when further inquired, he realized that he did not actually know why, he only had picked up what everybody was saying. And when he finally read thoroughly Bohm’s original papers, he understood that nothing was wrong and motivated the students to use the computer to calculate the trajectories of particles using Bohm's model. This marks the revival of Bohm’s hidden variables (see also \cite{junior2019david} Ch. 6.1), a revival to whom Bohm, however, obviously did not participate. Actually, when approached  by Dewdney  Philippidis, “Bohm himself [...] admitted that he had made a tactical error in his original presentation of the theory. The term \textit{hidden variables}, he said, created the wrong impression, and the papers themselves were too rigid and deterministic." (\cite{peat1997infinite}, p. 266). 

%\textcolor{blue}{[A few sentence on the following decades?]. see Freire Ch. 5. Perhaps mention how his causality and nonmechanism fit into his new view on the wholeness and the implicate order?}

%\textcolor{red}{[I added the implicate order related parts here}

In the following decades Bohm dedicated his work to an holistic approach that continued his ideas from the work on the causal interpretation of quantum theory. The purpose of Bohm’s original proposal in the light of his new ideas was later explained by himself in the following way: 

\begin{quote}
To show that it was wrong to throw out hidden variables
because they could not be imagined, it was therefore sufficient
to propose any logically consistent theory that explained the
quantum mechanics, through hidden variables, no matter how
abstract and hypothetical it might be. Thus, the existence of even
a single consistent theory of this kind showed that whatever
arguments one might continue to use against hidden variables,
one could no longer use the argument that they are inconceivable.
Of course, the specific theory that was proposed was not
satisfactory for general physical reasons, but if one such theory is
possible, then other and better theories may also be possible, and
the natural implication of this argument is ‘Why not try to find
them?’ (\cite{bohm1980wholeness}, p. 104)
\end{quote}

His scientific program was based on quantum field theory to approach the concept of the infinite levels he already pointed out in the early works. His philosophical ideas remained consistent to his early works in the refusal of mechanistic ideas: 

\begin{quote}
As we have seen, relativity theory requires continuity, strict causality (or determinism) and
locality. On the other hand, quantum theory requires noncontinuity,
non-causality and non-locality. So the basic concepts
of relativity and quantum theory directly contradict each other.
[...]
%It is therefore hardly surprising that these two theories have
%never been unified in a consistent way. Rather, it seems most
%likely that such a unification is not actually possible. 

What is very probably needed instead is a qualitatively new theory, from
which both relativity and quantum theory are to be derived as
abstractions, approximations and limiting cases.
The basic notions of this new theory evidently cannot be
found by beginning with those features in which relativity and
quantum theory stand in direct contradiction. The best place to
begin is with what they have basically in common. This is
undivided wholeness. Though each comes to such wholeness in
a different way, it is clear that it is this to which they are both
fundamentally pointing.
To begin with undivided wholeness means, however, that we must drop the mechanistic order. (\cite{bohm1980wholeness}, p. 223)   
\end{quote}

%he thought philosophers were narrow-minded in their appreciation of his major scientific-philosophical work, the book Causality and Chance in Modern Physics.”

%%%%%%%%%%%%%%%%%%%%%%%%%%%%%%%%%%%%%%%%%%%%%%%%%%%%%%%%%%%%%%%%%%%%%%%%%%%%%%%%%%%%%%%%%%
\subsection{Propensities and the causal interpretation}
\label{potential}
Bohm has been in touch with Popper at least since 1959 (for the relationship between them, see \cite{del2019karl} and references therein). It is exactly in that period that Popper---who was advocating for fundamental indeterminism in physics even at the classical level---developed a new interpretation of probabilities that are interpreted as objective physical properties, i.e., propensities or tendencies for a system to produce an outcome \cite{popper1959propensity}. 

Here we would like to stress that although Bohm's never actually pursued a program based on potentialities, he hinted at it in several occasions (see above). As we have seen, he endorsed that view in his Quantum Theory \cite{bohm1951quantum} and hinted that the statistical behaviors of quantum mechanics constrains the tendency of the sub-quantum fluid in his paper with Vigier \cite{bohm1954model}. Looking at Bohm’s correspondence with Popper, we find an explicit support of this view:    ``I feel that what you have to say about propensities make a genuine contribution to clarifying the issue that you discuss" (Bohm to K. Popper on March 15th 1967. PA, Popper’s Archives, Box/Folder: 84/19. AAU, Klagenfurt (Austria)/Hoover Institution, Stanford (California) \cite{Bohmletter84/19toPopper}.
This was not appreciated by Popper himself, who should be listed among the many that misinterpreted Bohm, attributing to him a strong commitment to determinism. In fact, when Popper published his book on the foundations of quantum theory in 1982 \cite{popper1982vol}, although prizing Bohm for striving for realism, he harshly criticized him for being a determinist. Bohm replied to him, emphasizing once again that he was not committed to determinism and explicitly acknowledging for the first time, to our knowledge, that his view on the causal interpretation can be regarded in terms of potentialities:
\begin{quote}
``I certainly think that a realistic interpretation of physics is essential. I think also that I understand your propensity interpretation of probability and I have no objections against it. […]. However, I feel that you have not properly understood my own point of view, which is much less different from yours than is implied in your book. Firstly I am not wedded to determinism. It is true that I first used a deterministic version of […] quantum theory. But later, with Vigier, a paper was written, in which we assumed that the movement of the particle was a stochastic process. Clearly that is not determinism. Indeed, we can regard the stochastic movement of the particle as affected by a field of propensities, in accordance with your ideas […] The key question at issue is therefore not that of determinism vs. indeterminism. I personally do not feel addicted to determinism [...].

[W]hat is real has a being independent of the consciousness of the observer. John Bell has used the term ``beable" to describe such an independent reality. From the point of view of realism, the main criticism of the orthodox interpretation of the quantum theory is that it has no room in it for ``beables". [...] I introduced the notion that the ``beables" of the quantum theory are the particles and the wavefunction (which contains information about the propensities). Along with Vigier, I can say that the ``beables" are themselves conditioned by such propensities. What are called the observables of  quantum theory  are then potentialities of the ``beables", realized according to a context, which in current physics, is determined by the experimental arrangement (though in nature, similar contexts will still exist without the intervention of human being). [...] My proposal has been that the ``beables" are particles (moving stochastically), along with the wave function. (Bohm to K. Popper 13.07.1984. Box/Folder: 278/2. AAU, Klagenfurt (Austria)/Hoover Institution, Stanford (California) \cite{Bohmletter278/2toPopper})  
\end{quote}

%%%%%%%%%%%%%%%%%%%%%%%%%%%%%%%%%%%%%%%%%%%%%%%%%%%%%%%%%%%%%%%%%%%%%%%%%%%%%%%%%%%%%%%%%%
\section{Discussion and conclusion}

In this paper, we have shown that Bohm was always against mechanism and therefore determinism. We have rebutted the historical narrative according to which one can identify an early period when Bohm was a supporter of Bohr, a later period when he was a committed determinist (influenced by Einstein and by Marxism), and finally a period, after his break with Marxism, in which determinism quit being a main concern of his. On the contrary, Bohm's philosophical tenets have never changed throughout his whole life: he was always committed to develop a realistic, causal, non-mechanistic view of physics. This led him to develop a new dialectical philosophy composed of infinite levels of description that guided him in his work for the following decades. As such, Bohm would have never accepted determinism, at any stage of his life. In a slogan, \textit{Bohm was never a Bohmian}.
%A clarification is here in order. 

Although the content of this paper has mostly a historical scope, it may affect also the physicists and philosophers who have proclaimed themselves Bohmians. It is undeniably true that Bohm provided the first deterministic hidden variable model of quantum theory. And yet, we just want to stress that for him this was nothing more than a model, a proof of principle that it was possible to do what was considered fundamentally unattainable. 
However, at the same time, this was for him most unsatisfactory, for it betrayed one of his deepest convictions about nature, namely, that a basic ontology of particles moved around by deterministic laws cannot be the end of the story. Therefore, the many scholars who today support Bohmian mechanics at face value, giving to it  an ontological role, should be aware that they are advocating a worldview that stems from what its original proposer considered a mere model which could not satisfy the basic standards of acceptability for a physical theory (except internal consistency). Now, while this is obviously a logically acceptable position, they should be aware that they are going directly against the fundamental views of Bohm, and  cannot therefore whatsoever appeal to his authority. This separation between the original though of Bohm and those who adopted his model was so striking that soon before his death when he became aware of Sheldon Goldstein and Detlev D\"urr's work on his ideas, Bohm bitterly confessed to his main collaborator Basil Hiley: ``why on earth are they calling it Bohmian mechanics? Haven't they read a word I have written?"     \cite{DelSantoKrizek2019}. So, concerning determinism, Bohm finds himself in a position comparable (fortunately with less ethical implications) to that Einstein with respect to the atomic bomb: It is a historical fact that it was Einstein who suggested to US president Franklin Roosevelt to research on nuclear weapons to preempt Nazi Germany to achieve the same threat. However, for  his whole life---before and after---Einstein was a committed pacifist. Similarly, it is a historical fact that Bohm developed a deterministic interpretation of quantum theory. However, for  his whole life---before and after---he was a committed anti-determinist. Invoking Bohm to defend deterministic views of physics is like invoking Einstein to promote nuclear weapons.

\subsection*{Acknowledgements}
%\textcolor{blue}{Basil Hiley, person from he Archive at Birkbeck College}
%\textcolor{red}{Please check if you are happy with that - I didn´t want to make it to cordially and personally in the paper. I would write Hiley an email in the aftermath}

The authors would like to thank Basil Hiley for taking time for an interview and valuable discussions. We also would like to express our thanks to Emma Illingworth from the David Bohm Archive at Birbeck Library for her support during our research. 

\bibliography{main}

\clearpage

%%%%%%%%%%%%%%%%%%%%%%%%%%%%%%%%%%%%%%%%%%%%%%%%%%%%%%%%%%%%%%%%%%%%%%%%%%%%%%%%%%%%%%%%%%
%%%%%%%%%%%%%%%%%%%%%%%%%%%%%%%%%%%%%%%%%%%%%%%%%%%%%%%%%%%%%%%%%%%%%%%%%%%%%%%%%%%%%%%%%%
%%%%%%%%%%%%%%%%%%%%%%%%%%%%%%%%%%%%%%%%%%%%%%%%%%%%%%%%%%%%%%%%%%%%%%%%%%%%%%%%%%%%%%%%%%
%%%%%%%%%%%%%%%%%%%%%%%%%%%%%%%%%%%%%%%%%%%%%%%%%%%%%%%%%%%%%%%%%%%%%%%%%%%%%%%%%%%%%%%%%%
%%%%%%%%%%%%%%%%%%%%%%%%%%%%%%%%%%%%%%%%%%%%%%%%%%%%%%%%%%%%%%%%%%%%%%%%%%%%%%%%%%%%%%%%%%
%%%%%%%%%%%%%%%%%%%%%%%%%%%%%%%%%%%%%%%%%%%%%%%%%%%%%%%%%%%%%%%%%%%%%%%%%%%%%%%%%%%%%%%%%%
%%%%%%%%%%%%%%%%%%%%%%%%%%%%%%%%%%%%%%%%%%%%%%%%%%%%%%%%%%%%%%%%%%%%%%%%%%%%%%%%%%%%%%%%%
%%%%%%%%%%%%%%%%%%%%%%%%%%%%%%%%%%%%%%%%%%%%%%%%%%%%%%%%%%%%%%%%%%%%%%%%%%%%%%%%%%%%%%%%%%

\section*{Appendix A -- Excerpts from correspondences of D. Bohm}

\subsection{Excerpt of a letter from Bohm to Miriam Yevick (January 7, 1952)}
\label{yevick07.01.52}
Letter 65. Folder C117, dated: Jan 7, 1952, \cite{talbot2017david}, p. 227.

    Now, to retain the concept of matter, we must above all retain the idea that
in some aspects at least, matter is indestructible and uncreatable. How then do we
explain the prevalence of change and the transiency of material things? This is done
by the notion of endless transformation. The “things” at each level, are made up of
smaller “elements” at a more fundamental level, and it is the  motion of these more fundamental
elements (not usually directly visible to us, except with the aid of elaborate
scientific research) which causes the appearance and disappearance of the “things”
existing at a higher level. These more fundamental “elements” however, cannot be
permanent, but must be made up of still more fundamental “elements” and so on ad
infinitum. Thus, we can see that every “thing” that exists may at some time come into
existence and later go out of existence, but there is always a deeper level, in terms of
which this change can be viewed rationally as a transformation of a more elementary
form of matter, which is not itself basically altered in this particular transformation.
Nevertheless, no single “thing” is uncreatable or indestructible. Only matter as a
whole in its infinity of properties and potentialities is eternal.

\subsection{Excerpt of a letter from Bohm to Schatzman; (not dated, 1952)}
 Letter A1.20, not dated, 1952. \cite{besson2018interpretation}, p. 351.
\label{schatzman1}

For quantum mechanics has show, that "empty" space a strongly fluctuating electromagnetic
fields and more important still, a very high density ( infinite according to the
present inadequate theories) of negative energy electrons, protons and neutrons. If
one adopts the new interpretation of the quantum mechanics, there is no choice but
co suppose chat these particles are really in existence. One therefore has been back to
the old notion of a material substratum filling all space. As a have said, this substratum
is very dense, much denser than any other form of matter. In fact, matter as it is
usually called, would be only a disturbance in the uniform background of substratum.
Light waves, etc. would also be disturbances of the substratum. The mysterious "annihilation"
 and "creation" of material particles could now be understood naturally;
for with the [ ?] of energy, the substratum could be made non-uniform as a spreading
wave. These two forms of energy could be transformed into each other when we look
out at the sky, space appears to be almost empty, because light waves are scattered only
by inhomogeneities in space. Similarly material particles are likewise inhomogeneities
propagated freely in a uniform background. Thus, to a naive way of looking, space appears
empty, a similar phenomenon appears in connection with the theory of metals.
As you know, an electron will go through a very dense metal without being scattered as 
long as the crystal lattice is perfectly regular. Only non-uniformities in the lattice will
scatter the electron. A naive observer (for example a positivist) would conclude from
this evidence that a metal consists of empty space, with a very thin haze of "matter" .
I would like to add one point here. It is most likely that not even the substratum
particles could be indestructible and unanalysable. Instead, there is probably another
substratum below this ( of a qualitatively different kind most probably) and so on ad
infinitum. Thus, we should have an infinite series of qualitatively different levels of
laws. Any finite number of levels can always be understood by humanity, but never
\underline{all} of them. Thus, ·we can understand more vividly a number of dialectical principles,
for example, many people are puzzled by the dialectical assertion that matter must be
eternal ( i.e. no creation). The answer is that at any particular level, the forms of matter
as a whole, in its infinite number of properties and inter -connections is eternal. Secondly,
consider the statement of dialectics chat  "a thing is not equal to itself" . this
we understand by the [ ? ] that a materiel "thing" contains an infinity of properties
whereas the concepts usually defining what the thing "is" cover only a finite number
of these properties. Thus, a thing is not only "what it is" but also a large nun1ber of
other things, which will manifest themselves later ; or in other words in "what is coming
to be". Moreover, the levels not taken into account in the usual definition of the
"theory" will generally produce effects that are in contradiction with the permanent
existence of this "thing" .

\subsection{Excerpt of a letter from Bohm to Miriam Yevick (January 23, 1952)}
\label{yevick23.01.52}
Letter 66. Folder C117, dated: Jan 23, 1952, \cite{talbot2017david}, p. 235:

[I]t is essential to think that things are not only “what they are known to be”, but also a
whole list of different things connected with the infinite number of levels not known
to us. These other things may be thought of roughly as “what is coming into being”
since it is in the future form of the thing that the underlying factors will ultimately
manifest themselves. [...]

%people see in change only a re-inforcement of the idea that “things are what they are”, but in order to hold on to this idea, they are forced to imagine that both external and human nature possess a terrifying and utterly irrational instability, because they are trying to explain what happens only in terms \underline{of what they can see now}. In order to avoid this terror, it is necessary to understand the levels underlying the visible level, to know the factors responsible for these changes, and thus to bring them under rational control. [...]

As in the structure of “elementary” forms of matter human beings contain an infinite number of at present unknown (or poorly known) levels of complexity of behavior.
This fact has two important implications: (1) The most obvious, that by scientific
study, we may ultimately learn to control some of the factors at any particular level,
and thus to produce startling changes in human nature (including even ourselves) (2)
Before this can be done, the different levels will manifest themselves in that people
cannot correctly be regarded as “being only what they are”, but that they can also
undergo fundamental transformations of character with changing conditions. [...]

As for the book [\cite{bohm1951quantum}], you must try to imagine the situation when I wrote it. You
suggest that I may have had some dishonesty, perhaps some desire to please the
“big shots” in writing it, and that this led me to back up the usual interpretation of
the quantum theory. You must remember several things however: (1) When I wrote
this book, there did not exist anywhere a clear statement of the basis of the theory.
There existed some books which made ridiculous abstract mathematical postulates that no one could possibly understand, and there were other discussions, such as those of Bohr, which aimed at discussing the physics, but in an incredibly vague way. A student at Princeton once told me that Bohr’s statements not only cancelled out with regard to their meaning in the first order, but also with regard to connotation in the second order. It was therefore necessary to go to the third order to find what Bohr meant. When I first started to study this subject 15 years ago, it fascinated me and puzzled me. I had no reason to suspect that the “big shots” had muddled up
the subject, since after all, had they not been astonishingly successful in predicting experiment after experiment? Above all, I never got over being puzzled by the theory.
When I started the book, I was in no position to see through the matter, because I still hadn’t made complete sense of it. All I knew was that there was one school, which utterly repelled me, in which one was supposed to introduce abstract mathematical postulates, and be satisfied if the calculations agreed with experiment. Against this,
Bohr’s school seemed to be a big improvement, because at least he tried to explain the physical meaning of the theory. Moreover, there was an element of dialectics in Bohr’s
point of view which attracted me. It seemed progressive because it broke the old
mechanist materialist determinism, which left no room for growth and development
of something new. After I had written the book, I finally began to grasp the full
meaning of the theory, and could see that it leads inevitably to a form of (dialectical)
idealism. But this was not so clear when I started, because of the general confusion
in the literature. If you tried to read other books, you wouldn’t be able to say that you
see through this stuff, just because the other books leave things just vague enough
so that you don’t know quite what you are seeing through. In writing this book,
I hope that I have not only clarified the issues for myself, but perhaps for other
people too. I suspect that a clear presentation of Bohr’s point of view (the first clear one, if I may boast a little) will do more to favor the causal interpretation than to favor Bohr’s interpretation. Now with my new point of view, I can see an infinitely
better way to get out of the trap of mechanistic determinism; namely through the
concept of an unlimited number of causal levels. I would call Bohr’s point of view
“static dialectics”. This is because it is a form of “slinging the lingo” in which the
dialectically opposing concepts are made just vague enough so that the contradictions
between them are avoided. Thus, one is not faced with the necessity of seeking new
concepts that synthesise the opposites, and the dynamic aspects of dialectics (i.e.
the contradictions leading to something new at another level) are lost. Finally, I
should say that I wrote the book in a spirit of struggle against the obscurantist notion
that nature can from now on be understood only in terms of abstract mathematical
postulates. The struggle was well worth while, since it led me to a new point of view.

\subsection{Excerpt of a letter from Bohm to Miriam Yevick (March 31, 1952)}
\label{yevick31.03.52}
Letter 73. Folder C118, dated: Rec Mar 31 [1952], \cite{talbot2017david}, pp. 254-55:

 I think that the explicit recognition of a limitless
number of levels would be a big step forward in science. Most of the errors of both
the positivist and the 19th century “mechanical” materialists spring from an implicit
assumption that the laws of nature will some day finally be understood in terms
of a limited number of hypotheses. From this comes the nightmare of a mechanically
determined universe that follows an inevitable course. To avoid this nightmare,
positivists and idealists have given up causality and assumed a “spontaneous” (i.e.,
uncaused) element in physical processes. [...]

The concept of a limitless number of levels suggests, however
that the work of science is never finished and leads one at each level to seek the
contradictions which can [unreadable] at the next level etc. Thus it provides a motive
power for continual development \& growth. Moreover, the nightmare of complete
determinism is avoided. Although each level is causal, the totality of levels cannot
ever be taken into account. Thus, as a matter of principle, we say that complete determinism
could not even be conceived of, yet, each level can be determined. Here, we
part company with the believers in “spontaneity” for we say that what appears to be
spontaneous is caused by factors, in principle, knowable, but now hidden to us. But
to be able to say this without implying complete determinism, we must assume an
unlimited number of levels. It is the unlimited number of levels which give matter
its “non-mechanical” aspects, for if the analysis of physical laws could ever be completed,
the theory would either be deterministic + “mechanical”, or “indeterministic” and “spontaneous”. Another interesting point – if there are an infinite number of levels,
we can expect that all existing limitations (such as speed of light and uncertainty
principle) can be overcome with the aid of more fundamental levels. Thus, by the use
of causal laws, humanity can move toward freedom. Whereas, in the ignorance of
causal laws, humanity is enslaved either to determinism or to “spontaneity”, which,
being pure accident, is just as tyrannical.
One other point, a distinction between “determinism” and “causality”. Although
both words have roughly the same meaning, their implications are different. For
causality implies (a) that if you know the causes, you can predict the effects. (b)
That if you change the causes, you can change the effects in a predictable way.
But determinism implies only predictability. In fact, with complete determinism, it
would be impossible for us ever to change anything. Now, if there are a finite number
of levels, then complete causality obviously implies complete determinism. But if
there are an infinite number, then the two concepts part company. For we can have
complete causality at every level, in the sense that we can use this causality to change
the world in a predictable way,with the error in the predictions dependent only on our
level of knowledge; whereas we can in no sense conceive of the world as completely
determined. In this connection, note that the statement that new things can come
into existence is consistent with causality, only if what is already in existence has
an infinite number of levels. For if we have a finite number of causal levels, then
the future is already contained logically in the present, but not if we have an infinite
number. The appearance of qualitatively new things with time is possible with an
infinite number, because the effects of the limitless number of lower levels can always
surge up into a higher level (and vice versa) producing qualitative [missing words]
describable as a rearrangement of things already in existence.

\subsection{Excerpt of a letter from Bohm to Melba Phillips (October 13, 1953)}
\label{phillis13.10.53}
Letter 43. Folder C48, dated: Oct 13, 1953, \cite{talbot2017david}, p. 164:

Also an important additional aspect of causality needs to be discussed in more detail –
namely – causality as a means of determining the mode of being of qualitatively
new things, which grow out of the old things. The basic aspect of mechanism is that
(as in an idealized machine) the universe is conceived of as made of basic elements
(particles, fields, or what have you) which simply interact according to fixed roles, and
which themselves never change as a result of the processes in which they take part.
Naturally, every physical theory has some non-mechanistic aspects. For example, in
the field theory, new entities (waves+particle --- like singularities) can arise out of the
interconnections of the basic field elements through the field equations (especially
if the latter are non-linear). Also in a particle theory, new entities can arise out of
interactions. [...] Nevertheless, the basic elements in such theories are usually
conceived of as fixed and eternal. However, the concept of the infinity of levels shows
that there need exist in nature no such thing as a basic element which never changes.
Thus, causal laws not only determine the future in a mechanical sense; i.e., in the
sense of determining quantitative changes in the arrangements of entities whose
intrinsic character is fixed. The causal laws also tell when qualitative changes will
occur and may define the characteristics of the new entities that can come into being.
Thus, causality is a broader concept than that of mechanical determinism. It contains
limited mechanical determinism as a special case. Indeed, the concept of causality
is continually evolving with the development of science and other aspects of human
activity, so that the potential richness of this concept has no limit. In other words, we
may expect future generations to discover more and more aspects of the concept of
causality, thus transforming this concept in a way that we have at present no inkling
of. Yet these changes will not be arbitrary, but will instead grow in a definite way out
of the efforts to solve real problems presented by the successive levels of reality that
we shall be able to reach. A “mechanistic” attitude toward science however, tends
to limit the growth of our concepts in an arbitrary and dogmatically conceived way.
Such a mechanistic attitude refers not only, however, to the mechanistic determinists,
but also to the “mechanistic indeterminists”, who insist that in the quantum of action,
we have reached an ultimate, indivisible, and unanalyzable entity, which will never
be found to have a structure understandable in terms of a deeper level. In fact, the
quantum of action presents many aspects of the ultimate particles of the atomists,
so that the insistence that the quantum will never be analyzed is as mechanistic as a
theory of point particles following determined orbits. Similarly, the insistence that
chance+probability are not subject to a causal analysis at a deeper level constitutes a
mechanistic attitude toward these things, since chance+probability are conceived of
as existing in themselves and functioning under all possible circumstances according
to fixed rules. [...]

According to the mechanistic
indeterminists, it is fixed by an equally mechanical “chance” which is conceived
of as absolute and not itself capable of change or development. We may make an
analogy of a man who is offered the possibility of 100 different ways of being
executed. The deterministic school of executioners would choose the way according
to certain definite factors, e.g., the chemical concentration of the blood, the wave
- length of the light emitted from his skin, etc. The indeterministic school would
chose the way by spinning a roulette wheel. The non-mechanistic school would seek a qualitative change - i.e., to find a way to escape execution, taking advantage of all
factors, both “determinate” and “chance”. So the essential point is that because of
the infinite complexity and depth of the laws governing the nature of matter, no preassigned
scheme of things can remain adequate forever, not even if it is restricted
to being a general framework or outline. But this is just what most people find
it difficult to accept – perhaps because our society requires us to accept the idea
that a certain general form of social organization is inevitable, although within this
general framework, we may make various quantitative changes, either by chance, or
by determinate rule, as we please, as long as nothing essential is ever changed. [...]

My own opinion is that the
synthesis will eventually have to be on a still deeper level and will have to introduce
new kinds of entities that are neither particles nor fields, of which we have only a
vague idea at present.

\subsection{Excerpt of a letter from Bohm to Melba Phillips (March 15, 1954)}
\label{phillis15.03.54}
Letter 46. Folder C48, dated: March 15, 1954, \cite{talbot2017david}, p. 170:

    First of all, it is necessary to
sharpen the distinction between causality and mechanism (or deterministic mechanism).
Mechanism is characterized by two fundamental aspects:

(1) Everything is made of certain basic elements which themselves never change
in essence (i.e., qualitatively).

(2)All that these elements can do is to undergo some quantitative change according
to some fixed laws of change. For example, if they are bodies, they can move in space.
If they are fields, they can change their numerical values, etc. But the basic elements
themselves never undergo qualitative change.

If we postulate an infinity of levels, then we make a step beyond mechanism. For
the elements existing at each level are made of still smaller elements in motion (i.e.,
changing quantitatively), and the mode of being of the higher level elements arises
out of the motions of the lower level elements. Thus, there are no elements that can
never change.

Indeed, even if we have a finite number of levels, some qualitative change is
possible within a mechanistic theory. For example, with atoms in chaotic motion, we
obtain new large scale properties, such as pressure, temperature, etc., new entities,
such as gas, liquid, solid, and qualitative changes between them. Now, at first sight, it
may seem that we could eliminate the large-scale level by analyzing it in terms of its
basic molecular motions. And if there were a finite number of levels, this would be
true. But if there are an infinite number, then each level stands on a footing that is, in
the long run, as basic as that of any other. For every level has below it a deeper one.
Indeed, matter can be regarded as made up of the totality of all levels. Each level
makes its own specific contribution to the totality. Of course, each level finds an
image in others, so that one can deduce many properties of a given level by studying
other levels. Yet, there may be properties that cannot so be deduced. Not only may
these properties be peculiar to a given level, but they may involve “crossing” of levels. [...]

Now, a mechanical law is characterized by the fact that it specifies a rule governing
quantitative changes of elements that are fixed in nature. A more general causal
law may express the conditions governing qualitative change. But if it does this, it
must do something else that a mechanical law is never called upon to do. It must not
only determine the mode of change, but also the mode of being of the elements when
they are not changing. A mechanical law simply postulates a certain fixed and eternal
mode of being of the elements, so that there is a sharp separation between the laws of
change and the mode of being of the elements. A more general causal law does not
make such a sharp separation. Thus, in the theory of evolution, the principle of natural
selection enables us to say something about the mode of being of the various forms of
life, in terms of their past history of evolution, struggle for survival, etc. Similarly, in
embryology, one can in part, understand the characteristic properties of an animal at
a given stage of development in terms of its past history which helped make it what it
now is. Thus, a more general causal law may be historical in form. By this, I mean that
\ul{the very mode of being of the elements which enter into the laws is 
a  necessary consequence of the causal laws governing the whole chain of development.}[...]

A causal law may express the necessity of a fundamental qualitative change, so
that what develops may have something new in it. This something \underline{new} arise[s] as
a necessary consequence of what is old, and yet it is not just a rearrangement or
a quantitative change of the old elements.

\subsection{Excerpt of a letter from Bohm to Miriam Yevick (September 10, 1954)}
\label{yevick10.09.54}
Letter 121. Folder C124, dated: Sept 10 1954, \cite{talbot2017david}, p. 419-22:

    The picture which I propose is this: The totality of causal laws includes both
statistical and individual laws. We start with this totality as our basic reality. Then,
we may take various views of this totality, some of which stress the individual aspect
of the laws, and some of which stress the statistical aspect. But there is no such thing
as a perfect individual law, because there are always fluctuations and errors coming
from what has been left out. [...]

We start with the idea of a real world, which
is in a continual process of change and development. We must now find means of
analyzing this change and development. To begin, we seek those aspects that have a
relative permanence. Over a short period of time, these aspects may be idealized and
abstracted as having a being, conceived of as static. But like the mathematical point,
the notion of a property or an aspect of things as having such a static and complete
being is only a simplifying abstraction. In reality it does not have such static being,
as is shown by the fact that it changes after some time. The fundamental reality is that of matter in being and in process of change, or of becoming, as it may more
accurately be called. [...]

We note that causal laws are relationships
between various aspects of reality at different times. Depending on which aspects that
we find are necessary, possible, or convenient to relate, we will have different kinds
of causal laws, some more nearly statistical and some more nearly individual. But the
essential point is that one and the same system simultaneously obeys individual and
statistical laws. [...] Thus, we do not regard the world as made of certain fixed eternal basic elements,
satisfying corresponding laws. [...]

[S]tatistical laws are not \underline{purely} a matter of convenience and practicability. Moreover
every level of individual law ultimately has some deeper statistical basis. A more
accurate statement of the problem is thus:
Both for reasons of practical convenience and for reasons of principle, we study
statistical aggregates in their own right. [...]

What must be stressed however is that
individual and statistical laws are abstractions as limiting cases of laws in general, and that there remains before us the problem of formulating more general types
of laws that could connect these two limiting cases in a continuous and rationally
understandable way.

\section*{Appendix B -- Excerpts from the writings of D. Bohm}

\subsection{Excerpts from \textit{Causality and Chance} (1957)}
\label{causality1}
Evidently, then, the applicability of the theory of probability to scientific and other statistical problems
has no essential relationship either to our knowledge or to our ignorance. Rather, it depends only on the
objective existence of certain regularities that are characteristic of the systems and processes under
discussion, regularities which imply that the long run or average behaviour in a large aggregate of objects or
events is approximately independent of the precise details that determine exactly what will happen in each
individual case.
On the basis of the above considerations, we are then led to interpret the probability of, for example, a
given result in the game of dice as an objective property associated with the dice that are being used and
with the process by which they are thrown, a property that can be defined independently of the question of
whether or not we know enough to predict what will happen in each individual throw. (p. 27)

When we study any particular set of processes within one of its relatively autonomous contexts, we
discover that certain relationships remain constant under a wide range of changes of the detailed behaviour
of the things that enter into this context. Such constancy is interpreted not as a coincidence, but rather as an
objective necessity inherent in the nature of the things we are studying. These necessary relationships are
then manifestations of the causal laws applying in the context in question. These laws do not have to determine
a given effect uniquely. Instead, they may (in the case of one-to-many relationships) determine only that the
effect must remain within a certain range of possibilities. (p. 29)

Now, as we shall see in this chapter and in other parts of the book, the mechanistic philosophy has taken
many specific forms throughout the development of science. The most essential aspects of this philosophy
seem to the author, however, to be its assumption that the great diversity of things that appear in all of our
experience, every day as well as scientific, can all be reduced completely and perfectly to nothing more than
consequences of the operation of an absolute and final set of purely quantitative laws determining the
behaviour of a few kinds of basic entities or variables. (p. 37)

The essential change brought in by this new point of view was the introduction of an element of
arbitrariness into the theory. One still thought of the universe as a gigantic mechanical system with the
property that everything in it can in principle be reduced completely and perfectly to nothing more than the
results of purely quantitative changes taking place in suitable mechanical parameters. But instead of having
its behaviour determined completely in terms of definite laws governing these parameters, this universal
system could continually be subject to irregular alterations in the course of its motion. [...]

For we now see that there is a \textit{whole level} in which
chance fluctuations are an inseparable part of the mode of being of things, so that they must be interwoven
into the fabric of the theory of this level in a fundamental way. Thus, we have been led to take an important
step beyond the classical notion of chance as nothing more than the effects of contingencies that modify the
boundary conditions or introduce randomly fluctuating external forces in a way that is not predictable within the context of interest, but which play no essential part in the formulation of the basic laws that apply within such a context.

If we stopped at this point, however, we should, as we have seen in the previous chapter, merely have
switched from deterministic to indeterministic mechanism. To avoid indeterministic mechanism, we must
suppose that, in their turn, the chance fluctuations come from something else. Since, as Heisenberg and Bohr
have shown so well, there is no room in the quantum domain for anything to exist in which these
fluctuations might originate, it is clear that to find their origin we must go to some new domain. [...]

Of course, if
one were now to make the assumption that these new laws would surely be nothing more than purely causal
laws, one would then fall back into deterministic mechanism, while the similar assumption that they were
surely nothing more than laws of probability would throw one back into indeterministic mechanism. On-the
other hand, we have in the proposals made in this chapter avoided both these dogmatic and arbitrary
extremes, since we have considered, as the situation demanded, the possibility that there are new features to
the causal laws (a “quantum force” not appearing at higher levels) as well as to the laws of chance (random
fluctuations originating in the sub-quantum mechanical level).
Of course, as we have indicated in Section 5, we do not regard our earlier proposals as providing a
completely satisfactory and definitive interpretation of the laws of the quantum domain. The basic reason is,
in a sense, that the fundamental concepts considered in the theory (waves and particles in interaction) are
still very probably too close to those applying in the classical domain to be appropriate to a completely new
domain such as that treated in the quantum theory. (pp. 126-127)

Actually, however, neither causal laws nor laws of chance
can ever be perfectly correct, because each inevitably leaves out some aspect of what is happening in
broader contexts. [...] Thus, we are led to regard these two kinds of laws as effectively furnishing different views of any
given natural process, such that at times we may need one view or the other to catch what is essential, while
at still other times, we may have to combine both views in an appropriate way. But we do not assume, as is
generally done in a mechanistic philosophy, that the whole of nature can eventually be treated completely
perfectly and unconditionally in terms of just one of these sides, so that the other will be seen to be
inessential, a mere shadow, that makes no fundamental contribution to our representation of nature as a whole. (p. 143)

\section*{Appendix C -- Excerpts from the secondary literature about D. Bohm}

\subsection{Excerpt from Freire, O. Jr, David Bohm: A life dedicated to understanding the quantum world}
\label{freire91}
Évry Schatzman, who was the intermediary for Bohm to contact Vigier, wrote to Bohm: “Any physical theory should be completely deterministic, because an affirmation of the dialectical materialism is that there is an objective reality and that this reality is cognizable, that we can built an image of that reality in our mind”. Schatzman was far from modest about the work which was being done by Bohm and Vigier, comparing it to Marx’s works: “We should be grateful to people like Vigier, like you, who have with tenacity devoted their efforts to the rebuilding of the quantum theory on its feet, just like the dialectic
of Hegel, which had to be put back on its feet!” However, if the Marxist background
was the cement, the collaboration between Bohm and Vigier blossomed in a fruitful
scientific collaboration. (\cite{junior2014quantum}, p. 91)

%%%%%%%%%%%%%%%%%%%%%%%%%%%%%%%%%%%%%%%%%%%%%%%%%%%%%%%%%%%%%%%%%%%%%%%%%%%%%%%%%%%%%%%%%%
%%%%%%%%%%%%%%%%%%%%%%%%%%%%%%%%%%%%%%%%%%%%%%%%%%%%%%%%%%%%%%%%%%%%%%%%%%%%%%%%%%%%%%%%%%
%%%%%%%%%%%%%%%%%%%%%%%%%%%%%%%%%%%%%%%%%%%%%%%%%%%%%%%%%%%%%%%%%%%%%%%%%%%%%%%%%%%%%%%%%%
%%%%%%%%%%%%%%%%%%%%%%%%%%%%%%%%%%%%%%%%%%%%%%%%%%%%%%%%%%%%%%%%%%%%%%%%%%%%%%%%%%%%%%%%%%
%%%%%%%%%%%%%%%%%%%%%%%%%%%%%%%%%%%%%%%%%%%%%%%%%%%%%%%%%%%%%%%%%%%%%%%%%%%%%%%%%%%%%%%%%%
%%%%%%%%%%%%%%%%%%%%%%%%%%%%%%%%%%%%%%%%%%%%%%%%%%%%%%%%%%%%%%%%%%%%%%%%%%%%%%%%%%%%%%%%%%
%%%%%%%%%%%%%%%%%%%%%%%%%%%%%%%%%%%%%%%%%%%%%%%%%%%%%%%%%%%%%%%%%%%%%%%%%%%%%%%%%%%%%%%%%
%%%%%%%%%%%%%%%%%%%%%%%%%%%%%%%%%%%%%%%%%%%%%%%%%%%%%%%%%%%%%%%%%%%%%%%%%%%%%%%%%%%%%%%%%%

\clearpage

%\begin{quote}
 %   It was always something he came back to. The infinite quality of nature was his idea. So, whenever you think you got you got to the end of the story,  there is always something deeper. But that was just a general philosophical idea.  
%\end{quote}

%This is very important: Bohm was aware of how even his supporters were actually completely unaware of his ideas and only grabbed his 1952 paper.
%\begin{quote}
%    When he saw [Sheldon] Goldstein and Detlev D\"urr down to his ideas, and called it ``Bohmian mechanics", he said to me, and it was just before he died, he said to me: ``why on earth are they calling it Bohmian mechanics? Haven't they read a word I have written?"  
%\end{quote}

%About Bohm's political views
%\begin{quote}
%Bohm was a Marxist in America. When he came here, his views where right of British Labor Party. [...] And that was a tension between Vigier and himself.  
%\end{quote}

%\begin{quote}
%If you read ``Causality and Chance" \cite{bohm1957causality}, you will find a severe criticism against his own theory.  
%\end{quote}
%Hiley says that therein there is a full chapter against mechanism.

\clearpage

\end{document}